# Atomic Coherence Assisted Multipartite Entanglement Generation with DELC Four-Wave Mixing

Yuliang Liu*, Jiajia Wei*, Mengqi Niu, Yixin Lin, Zhili Chen, Jin Yan, Binshuo Luo, Feng Li, Yin Cai[†], and Yanpeng Zhang[‡]

*Key Laboratory for Physical Electronics and Devices of the Ministry of Education, Shaanxi Key Lab of Information Photonic Technique, School of Electronic Science and Engineering, Xi'an Jiaotong University, Xi'an, Shaanxi 710049, China*

*The authors contribute equally
[†]caiyin@mail.xjtu.edu.cn
[‡]ypzhang@mail.xjtu.edu.cn

**Abstract:** Multipartite entanglement plays an important role in quantum information processing and quantum metrology. Here, the dressing-energy-level-cascaded (DELC) four-wave mixing (FWM) processes are proposed to generate all-optical controlled multipartite entanglement within a single device. The entanglement characteristics of the produced states of light are characterized by applying the Duan criterion and the positivity under partial transposition criterion. Moreover, by using an internal dressing field to modulate atomic coherence, multiple quantum coherent channels of FWM are simultaneously constructed, which result in a great extension of entanglement mode number and quantum information capacity. We find that the violation of the entanglement criteria inequalities is coherent-channel dependent, and the produced states can be directly modulated via atomic coherence. Our system can integrate the generation and modulation of the entangled states in one process. It may help provide a compact method for realizing large scale quantum networks.

## 1. Introduction

Multipartite entanglement has garnered a lot of attention because of its significance and potential applications in quantum information processing and quantum metrology [1-5]. A mature technology for generating multiphoton entangled states is applying multiple spontaneous parametric down-conversion processes (SPDC), which produce down-converted photons pairs in a second-order nonlinear crystal [6-10]. Continuous-variable multipartite entanglement can be implemented by using multiple optical parametric oscillators (OPOs) [11-17]. Besides, many alternative methods to produce multipartite entanglement are demonstrated, such as quantum frequency combs [18,19], spatial modes [20], temporal entanglement [21,22], etc. Hitherto, multipartite entanglement has been widely applied in quantum sensing [23-25], quantum computing [26-28], and constructing quantum networks [19,30].

In recent years, another effective method for preparing entanglement is using four-wave mixing (FWM) process of atomic media. Entangled light beams have been experimentally verified through a parametric amplified FWM (PA-FWM) process in high-gain atomic media [30-33]. The FWM process exhibits unique advantages. It has the nature of spatial multimode and the produced entangled light beams are spatially separated [34,35]. Also, due to the strong nonlinearity, the optical cavity is not required and therefore the experimental setup is simplified. With the feature of scalability and flexibility, many methods to prepare multipartite entanglement using FWM are theoretically proposed and experimentally demonstrated, including multi-pump [36,37], cascading atomic cells [38-41], etc. However, the effect of the atomic coherence, which is important for generating and modulating multipartite entanglement in atomic ensemble, has not been explored.

In atomic ensemble, the nonlinear susceptibility of FWM can be actively modulated based on atomic coherence. The induced dressing effect can be used to reshape the parametric gain profile of FWM and realize coherent control of the quantum entanglement. Moreover, via splitting atomic energy levels with the dressing effect, the frequency modes of the correlated photons can be extended to construct multiple coherent channels of FWM, and it results in hyper entanglement and energy-time entanglement [42-44]. Also, with the constructive interference among different transition probability amplitudes, the conversion efficiency of the FWM process may increase, and therefore the degree of squeezing states of light can be enhanced by modulating the internal states of a multilevel atomic system [45-47].

In this paper, multipartite entangled states of light are one-step produced from dressing-energy-level-cascaded FWM (DELC-FWM). This scheme employs a single device of hot rubidium atomic medium and exhibits  advantages such as simplified experimental devices, lower optical path losses, and fewer vacuum losses. We apply the Duan [48] and the positivity under partial transposition (PPT) [49,50] criteria to investigate the multipartite entanglement of the output beams. Furthermore, we introduce the dressing field to simultaneously construct multiple coherent channels of FWM, thereby realizing multimode entanglement and expanding the quantum information capacity. In our work, with atomic coherence, the generation and modulation of multipartite entanglement can be integrated in the process of the entangled states preparation. These results may be helpful for providing a compact way in multimode quantum secure communication, quantum computing, and quantum sensing, etc.

## 2. Theoretical Model of DELC-FWM Processes
### 2.1 Generation of Three-Mode Outputs

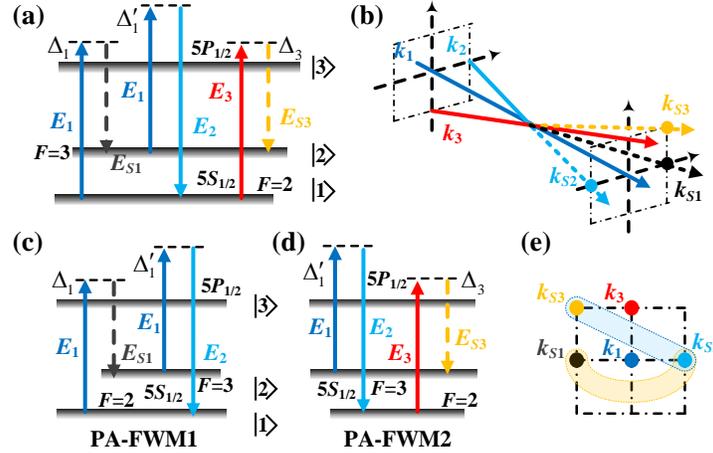

Figure 1. (a) Energy-level diagram of the three-mode DELC-FWM processes in the rubidium atomic system. (b) The spatial distribution of the beams. The arrows represent the signal beams. The angles are determined by the phase-matching conditions and the intersection point is the Rb cell. $\mathbf{k}_i$ is the wave vector of $E_i$ ($i$ = 1, 2, 3, $S$1, $S$2, $S$3). (c) Energy-level diagram of PA-FWM1 process. (d) Energy-level diagram of PA-FWM2 process. (e) The tangential distribution of output signal beams. The belts indicate quantum correlation existing between the two connected modes.

We consider a double-$\Lambda$ type three-level $|1\rangle$-$|2\rangle$-$|3\rangle$ atomic system, as shown in Figure 1(a). One possible experimental candidate for the proposed system is $5S_{1/2}$, $F$=2 ($|1\rangle$), $5S_{1/2}$, $F$=3 ($|2\rangle$) and $5P_{1/2}$, ($|3\rangle$) in $^{85}$Rb. A strong pump beam $E_1$ (frequency $\omega_1$, wave vector $\mathbf{k}_1$ and Rabi frequency $\Omega_1$) drives the transition $|1\rangle\rightarrow|3\rangle$ and $|2\rangle\rightarrow|3\rangle$ with frequency detuning $\Delta_1$ and $\Delta_1'$, respectively. A weak probe beam $E_2$ ($\omega_2$, $\mathbf{k}_2$ and $\Omega_2$) drives the transition $|3\rangle$ to $|1\rangle$. The pump field $E_3$ ($\omega_3$, $\mathbf{k}_3$ and $\Omega_3$) drives the transition $|1\rangle$ to $|3\rangle$ with frequency detuning $\Delta_3$. The entire process can be viewed as two PA-FWM processes (Figures 1(b) and 1(c)) cascaded together, which corresponds to cascading two Rb cells as shown in Figure A1 in Appendix. The detuning $\Delta_i$ is defined as the difference between the resonant transition frequency and the laser frequency of $E_i$. With the detuning of $E_1$ tuned far away from the resonance, the PA-FWM1 and PA-FWM2 will occur in the system, which can generate the quantum correlated output beams $E_{S1}$, $E_{S2}$ (amplified $E_2$) and $E_{S3}$ in a single Rb cell satisfying the phase-match conditions $\mathbf{k}_{S1} + \mathbf{k}_{S2} = 2\mathbf{k}_1$ and $\mathbf{k}_{S2} + \mathbf{k}_{S3} = \mathbf{k}_1 + \mathbf{k}_3$, respectively.

In this DELC-FWM system, three inputs $E_1$, $E_2$ and $E_3$ converge in the Rb cell and generate three spatially separated outputs $E_{S1}$, $E_{S2}$ (amplified $E_2$) and $E_{S3}$ by cascading two PA-FWM processes. The interaction Hamiltonian of PA-FWM1 process and PA-FWM2 process can be expressed as:

$$H_1 = i\hbar\kappa_1 \hat{a}_1^\dagger \hat{a}_2^\dagger + \text{H.c.}, \tag{1a}$$

$$H_2 = i\hbar\kappa_2 \hat{a}_2^\dagger \hat{a}_3^\dagger + \text{H.c.}, \tag{1b}$$

where $\hat{a}_1^\dagger$, $\hat{a}_2^\dagger$ and $\hat{a}_3^\dagger$ are the photon creation operators of the output modes of $E_{S1}$, $E_{S2}$ and $E_{S3}$, respectively; $\kappa_1 = -i\varpi_1\chi_1^{(3)}E_1^2/2c$ and $\kappa_2 = -i\varpi_2\chi_2^{(3)}E_1E_3/2c$ are the pumping parameter of PA-FWM1 and PA-FWM2 respectively, which depends on the nonlinear susceptibility $\chi_1^{(3)}$, $\chi_2^{(3)}$ and pump-field amplitude; $\varpi_i$ is the central frequency of generated signals; H.c. is Hermitian conjugate.

The boson-creation (-annihilation) operator satisfies the Heisenberg operator of motion in the dipole approximation. The dynamic equation of the system can be written as $\frac{d\hat{a}_i}{dt} = \frac{-i}{\hbar}[\hat{a}_i, H]$, ($i$=1,2,3), from which we obtain:

$$\text{PA-FWM1} \quad \frac{d\hat{a}_1}{dt} = \kappa_1 \hat{a}_2^\dagger, \quad \frac{d\hat{a}_2}{dt} = \kappa_1 \hat{a}_1^\dagger \tag{2a}$$

$$\text{PA-FWM2} \quad \frac{d\hat{a}_2}{dt} = \kappa_2 \hat{a}_3^\dagger, \quad \frac{d\hat{a}_3}{dt} = \kappa_2 \hat{a}_2^\dagger \tag{2b}$$

After the operation of time evolution equation, the final input-output relations of this system are:

$$\hat{a}_{1out} = G_1 \hat{a}_{1in} + g_1 \hat{a}_{2in}^\dagger \tag{3a}$$

$$\hat{a}_{2out} = g_1 G_2 \hat{a}_{1in}^\dagger + G_1 G_2 \hat{a}_{2in} + g_2 \hat{a}_{3in}^\dagger \tag{3b}$$

$$\hat{a}_{3out} = g_1 g_2 \hat{a}_{1in} + G_1 g_2 \hat{a}_{2in}^\dagger + G_2 \hat{a}_{3in} \tag{3c}$$

where $G_1 = \cosh(\kappa_1 t)$ and $G_2 = \cosh(\kappa_2 t)$ are the amplitude gain in the PA-FWM1 and PA-FWM2, respectively; $G_i^2 - g_i^2 = 1$ ($i$=1, 2). $t$ is the interaction time which can be adjusted by changing the length and temperature of the Rb vapor in experiment. $\hat{a}_{1in}^\dagger$, $\hat{a}_{2in}^\dagger$ and $\hat{a}_{3in}^\dagger$ ($\hat{a}_{1in}$, $\hat{a}_{2in}$ and $\hat{a}_{3in}$) are the creation (annihilation) operators of inputs $\boldsymbol{E}_1$, $\boldsymbol{E}_2$ and $\boldsymbol{E}_3$, respectively; $\hat{a}_{1out}$, $\hat{a}_{2out}$ and $\hat{a}_{3out}$ are the annihilation operators of the outputs $\boldsymbol{E}_{S1}$, $\boldsymbol{E}_{S2}$ and $\boldsymbol{E}_{S3}$, respectively. The amplitude and phase quadrature operators are defined as $\hat{X} = \hat{a} + \hat{a}^\dagger$ and $\hat{P} = i(\hat{a}^\dagger - \hat{a})$. A vector of canonical quadrature operators is defined as $\mathbf{r} = (\hat{X}_1, \hat{P}_1, \cdots, \hat{X}_n, \hat{P}_n)$. The evolution equations of the three output modes can be written as: $\mathbf{r}_{out} = U_{tri} \mathbf{r}_{in}$, where the transform operation matrix $U_{tri}$ are:

$$U_{tri} = \begin{pmatrix} G_1 & 0 & g_1 & 0 & 0 & 0 \\ 0 & G_1 & 0 & -g_1 & 0 & 0 \\ g_1 G_2 & 0 & G_1 G_2 & 0 & g_2 & 0 \\ 0 & -g_1 G_2 & 0 & G_1 G_2 & 0 & -g_2 \\ g_1 g_2 & 0 & G_1 g_2 & 0 & G_2 & 0 \\ 0 & g_1 g_2 & 0 & -G_1 g_2 & 0 & G_2 \end{pmatrix} \tag{4}$$

The quantum correlations of multimode Gaussian states generated by a quadratic Hamiltonian can be fully characterized by its covariance matrix (CM). The elements of the CM are defined as follows:

$$\sigma_{ij} = \frac{1}{2}\langle \hat{r}_i \hat{r}_j + \hat{r}_j \hat{r}_i \rangle - \langle \hat{r}_i \rangle \langle \hat{r}_j \rangle \tag{5}$$

According to the definition of the CM, we can reconstruct the CM as $\sigma = U_{tri} U_{tri}^T$, based on Equation 4 with the inputs of coherent or vacuum states. Note that there does not exist cross-correlation between the amplitude and phase quadrature in this system, and the CM is positive and symmetric.

## 2.2 Generation of Four-Mode Outputs

The relevant energy level to generate four-mode Gaussian states in the Rb atomic system is the same as that in the three-mode case but the pump field $\boldsymbol{E}_3$ ($\omega_3$, $\mathbf{k}_3$, and $\Omega_3$) drives the transitions $|1\rangle \rightarrow |3\rangle$ and $|2\rangle \rightarrow |3\rangle$ with frequency detuning $\Delta_3$ and $\Delta_3'$, respectively. This four-mode DELC-FWM system contains an eight-wave mixing process that satisfies the phase-match condition $2\mathbf{k}_1 + 2\mathbf{k}_3 = \mathbf{k}_{S1} + \mathbf{k}_2 + \mathbf{k}_{S3} + \mathbf{k}_{S4}$ and four FWM processes that satisfy $2\mathbf{k}_1 = \mathbf{k}_{S1} + \mathbf{k}_2$, $\mathbf{k}_1 + \mathbf{k}_3 = \mathbf{k}_2 + \mathbf{k}_{S3}$, $\mathbf{k}_1 + \mathbf{k}_3 = \mathbf{k}_{S1} + \mathbf{k}_{S4}$, and $2\mathbf{k}_3 = \mathbf{k}_{S3} + \mathbf{k}_{S4}$, respectively. One of the simplified method to obtain the input-output relations is to view the system as three PA-FWM processes cascaded, which corresponds to cascading three Rb cells as shown in Figure A2 in Appendix [51]. The outputs of PA-FWM1, amplified $\boldsymbol{E}_2$ and $\boldsymbol{E}_{v0}$, are injected into PA-FWM2 and PA-FWM3, respectively. The interaction Hamiltonian of these three PA-FWM processes can be expressed as:

$$H_1 = i\hbar \kappa_1 \hat{a}_1^\dagger \hat{a}_2^\dagger + \text{H.c.} \tag{6a}$$

$$H_2 = i\hbar \kappa_2 \hat{a}_2^\dagger \hat{a}_3^\dagger + \text{H.c.} \tag{6b}$$

$$H_3 = i\hbar \kappa_3 \hat{a}_1^\dagger \hat{a}_4^\dagger + \text{H.c.} \tag{6c}$$

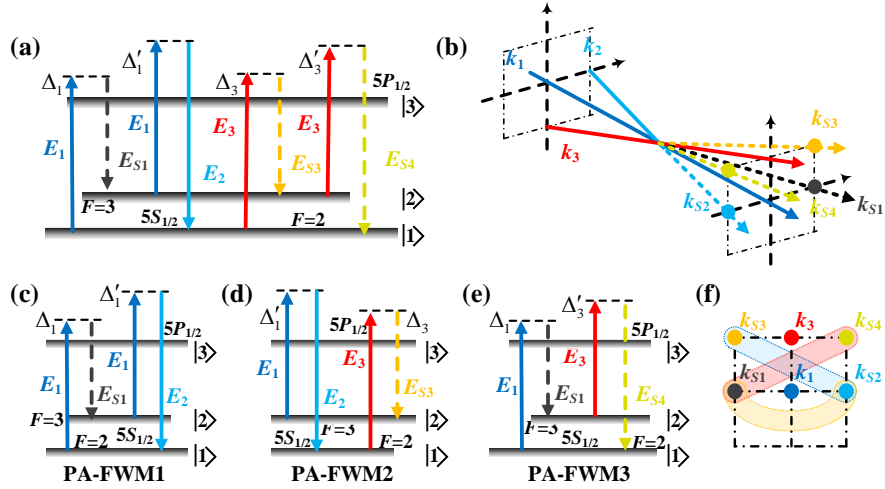

Figure 2. (a) Energy-level diagram of the four-mode DELC-FWM processes in the rubidium atomic system. (b) The spatial distribution of the beams. The arrows represent the signal beams. The angles are determined by the phase-matching conditions. (c)-(e) Step breakdown energy-level diagram of the subsystem. Three of the four PA-FWM processes can compose the four mode DELC-FWM processes. (f) The tangential distribution of output signal beams. The belts indicate quantum correlation existing between the two connected modes.

The amplitude gain and interaction strength are as follows:

$$G_1 = \cosh(\kappa_1 t), \kappa_1 = -i\varpi_1 \chi_1^{(3)} E_1^2 / 2c \tag{7a}$$

$$G_2 = \cosh(\kappa_2 t), \kappa_2 = -i\varpi_2 \chi_2^{(3)} E_1 E_3 / 2c \tag{7b}$$

$$G_3 = \cosh(\kappa_3 t), \kappa_3 = -i\varpi_3 \chi_3^{(3)} E_3^2 / 2c \tag{7c}$$

Using the similar method as the three-mode system, we can obtain the final input-output relations and then obtain the CM. The transform operation matrix $U_{quad}$ in this four-mode system can be written as:

$$U_{quad} = \begin{pmatrix} G_1G_3 & 0 & g_1G_3 & 0 & 0 & 0 & g_3 & 0 \\ 0 & G_1G_3 & 0 & -g_1G_3 & 0 & 0 & 0 & -g_3 \\ g_1G_2 & 0 & G_1G_2 & 0 & g_2 & 0 & 0 & 0 \\ 0 & -g_1G_2 & 0 & G_1G_2 & 0 & -g_2 & 0 & 0 \\ g_1g_2 & 0 & G_1g_2 & 0 & G_2 & 0 & 0 & 0 \\ 0 & g_1g_2 & 0 & -G_1g_2 & 0 & G_2 & 0 & 0 \\ G_1g_3 & 0 & g_1g_3 & 0 & 0 & 0 & G_3 & 0 \\ 0 & -G_1g_3 & 0 & g_1g_3 & 0 & 0 & 0 & G_3 \end{pmatrix} \tag{8}$$

## 3. Multipartite Entanglement

In this section, Duan and PPT criteria are used to investigate the multipartite entanglement of the generated quantum Gaussian states. The inequalities of Duan criterion using the quantum correlations of the associated modes are defined as follows:

$$D_{ij} = V(\hat{X}_i - \hat{X}_j) + V(\hat{Y}_i + \hat{Y}_j) \geq 4 \tag{9}$$

where $V(\hat{X}_i - \hat{X}_j)$ is the variances of the difference of the amplitude quadratures and $V(\hat{Y}_i + \hat{Y}_j)$ is the sum of the phase quadratures. The values $D_{ij}$ suggests the amount of the inseparability. The bipartite entanglement between mode $\hat{a}_i$ and $\hat{a}_j$ can be demonstrated by the violation of the inequalities.

PPT criterion can be used to characterize the entanglement of two subsystems which consists of one or several modes and the smaller symplectic eigenvalue suggests the inseparability. It is a sufficient and necessary criterion for the case of 1 versus $n$-modes (1-$n$) and only sufficient for the case of $m$-$n$. For bipartite entanglement of $\sigma_{AB}$, the partial transposition operation on part

A is equivalent to the transformation through matrix $T_A = \left( \oplus_{k=1}^{m} \text{diag}(1,-1) \right)_A \oplus I_B$, where the first factor is its mirror reflection in phase space, and the second factor represents the other subsystems. The criterion will lead to the following uncertainty: $\tilde{\sigma}_{AB} + i\Omega \geq 0$ where $\tilde{\sigma}_{AB} = T_A \sigma_{AB} T_A$, and $\Omega$ is symplectic form with same dimension of $\tilde{\sigma}_{AB}$. The existence of negative symplectic eigenvalues directly means the presence of entanglement.

## 3.1 Three-Mode Outputs

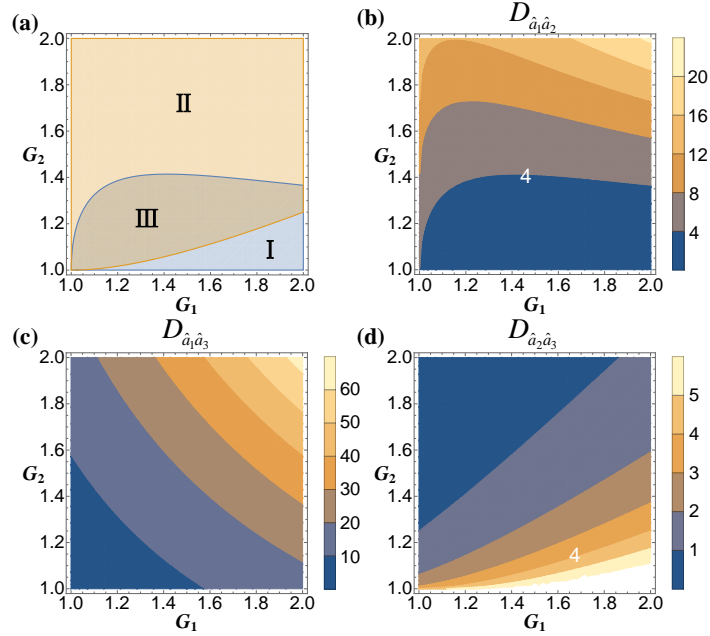

Figure 3. Bipartite entanglement for the three-mode DELC-FWM processes using Duan criterion. (a) Only $D_{12}$ is less than 4 in region I. Only $D_{23}$ is less than 4 in region II. $D_{12}$ and $D_{23}$ are both less than 4 in region III. (b)-(d) Duan values of $D_{12}$, $D_{13}$ and $D_{23}$ with different values of the gains $G_1$ and $G_2$.

Duan criterion can be applied to verify any two modes of the three produced modes $\hat{a}_1$, $\hat{a}_2$ and $\hat{a}_3$. In the three-mode system, the values of Duan criterion $D_{12}$, $D_{13}$ and $D_{23}$ are given by

$$D_{12} = 4\left( G_1^2 \left( G_2^2 + 1 \right) - 2G_1 G_2 \sqrt{G_1^2 - 1} - 1 \right) \tag{10a}$$

$$D_{13} = 4G_1^2 G_2^2 \tag{10b}$$

$$D_{23} = -4G_1^2 \left( 2G_2 \sqrt{G_2^2 - 1} - 2G_2^2 + 1 \right) \tag{10c}$$

The expression show the dependence of $D_{12}$, $D_{13}$ and $D_{23}$ on the gains $G_1$ and $G_2$. The region plots are shown in Figure 3. The amount of the inseparability $D_{ij}$ larger than 4 means no entanglement. The entanglement region of modes $\hat{a}_1$ and $\hat{a}_2$ ($D_{12} < 4$) is the blue region I and III in Figure 3(a). The entanglement between modes $\hat{a}_1$ and $\hat{a}_2$ is very sensitive to the $G_2$. They entangle when $G_2$ is smaller, because only beam $\hat{a}_2$ is amplified with the second PA-FWM process, which leads to their quantum noise unbalance. $D_{13}$ is larger than 4 for any $G_1 > 1$ and $G_2 > 1$ in Figure 3(c). This is due to modes $\hat{a}_1$ and $\hat{a}_3$ do not interact with each other in the Hamitonian in Equation 1, and thus there is never entanglement existing between modes $\hat{a}_1$ and $\hat{a}_3$. The entanglement region of modes $\hat{a}_2$ and $\hat{a}_3$ ($D_{23} < 4$) is the orange region II and III in Figure 3(a). The entanglement of modes $\hat{a}_2$ and $\hat{a}_3$ is limited by bigger $G_1$, that is, a bigger $G_1$ is not helpful in the presentence of the entanglement.

The PPT criterion is also verified in this three-mode system. PPT criterion can be used to characterize the bipartite and tripartite entanglement. The value of PPT criterion smaller than 0 directly means the presence of entanglement. The results of the bipartite entanglement characterized by PPT criterion are similar as Duan criterion. The PPT value of $\hat{a}_1$-$\hat{a}_3$ is larger than 0 for any $G_1 > 1$ and $G_2 > 1$, which is consistent with previous analysis. The entanglement exist in all 1-2 modes types. We find the trends of tripartite entanglement $\hat{a}_1$-$\{\hat{a}_2\hat{a}_3\}$ and $\hat{a}_3$-$\{\hat{a}_1\hat{a}_2\}$ are similar to the entanglement of $\hat{a}_1$-$\hat{a}_2$ and $\hat{a}_2$-$\hat{a}_3$ respectively, i.e., the conjugate beam $\hat{a}_1$ or $\hat{a}_3$ added with the probe beam $\hat{a}_2$ becoming $\{\hat{a}_1\hat{a}_2\}$ or $\{\hat{a}_2\hat{a}_3\}$ can only quantitatively change the amount of entanglement and do not change the dependence on parameters $G_1$ and $G_2$.

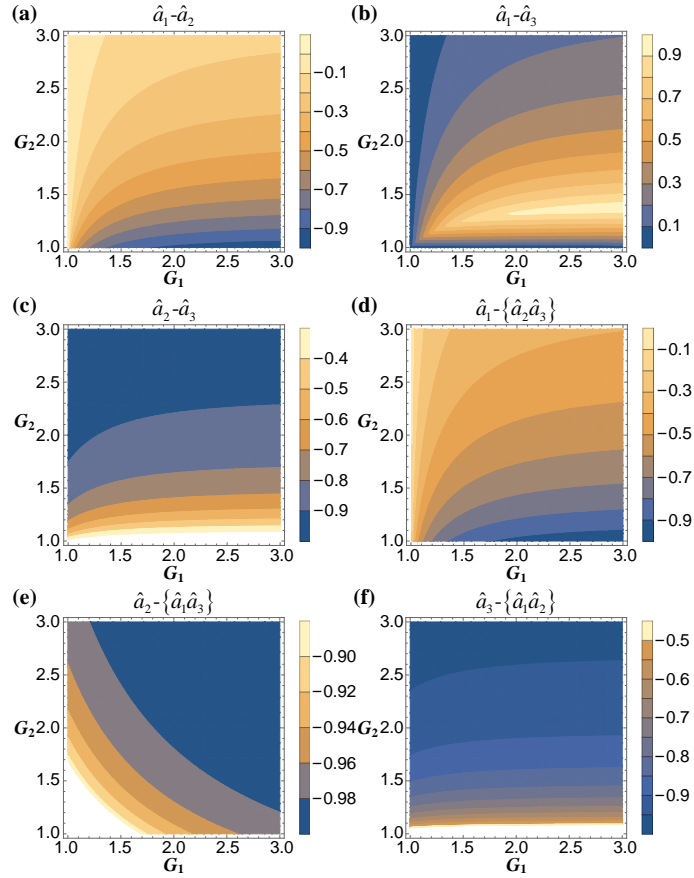

Figure 4. PPT values of the three-mode DELC-FWM processes. (a)-(c) and (d)-(f) are bipartite and tripartite entanglement region plots, respectively.

### 3.2 Four-Mode Outputs

In the four-mode system, all possible two-mode permutations of Duan criterion is verified, and the dependence of Duan values on the gains $G_1$, $G_2$ and $G_3$ are shown in Equation 11. The corresponding contour plots are shown in Figure 5 where we set the gain of the first PA-FWM process is 1.1.

$$D_{12} = 4\left(G_1^2 G_2^2 + G_1^2 G_3^2 - 2G_1 G_2 G_3 \sqrt{G_1^2 - 1} - 1\right) \tag{11a}$$

$$D_{13} = D_{24} = 4G_1^2 \left(G_2^2 + G_3^2 - 1\right) \tag{11b}$$

$$D_{14} = 4G_1^2 \left(2G_3^2 - 2G_3 \sqrt{G_3^2 - 1} - 1\right) \tag{11c}$$

$$D_{23} = 4G_1^2 \left(2G_2^2 - 2G_2 \sqrt{G_2^2 - 1} - 1\right) \tag{11d}$$

$$D_{34} = 4\begin{pmatrix} -2G_1^2 + G_1^2 G_2^2 + G_1^2 G_3^2 \\ -2G_1 \sqrt{G_1^2-1}\sqrt{G_2^2-1}\sqrt{G_3^2-1} + 1 \end{pmatrix} \tag{11e}$$

The values of $D_{12}$, $D_{13}$, $D_{24}$ and $D_{34}$ all go up with the increasing $G_2$ and $G_3$, suggesting that stronger $G_2$ and $G_3$ do not help in enhancing the entanglement for a fixed $G_1$. Moreover, $D_{14}$ ($D_{23}$) changes with the gain $G_1$ and $G_3$ ($G_1$ and $G_2$), because modes $\hat{a}_1$ and $\hat{a}_4$ (modes $\hat{a}_2$ and $\hat{a}_3$) never participate in the PA-FWM2 (PA-FWM3) process and therefore gain $G_2$ ($G_3$) does not contribute to the entanglement between these two modes.

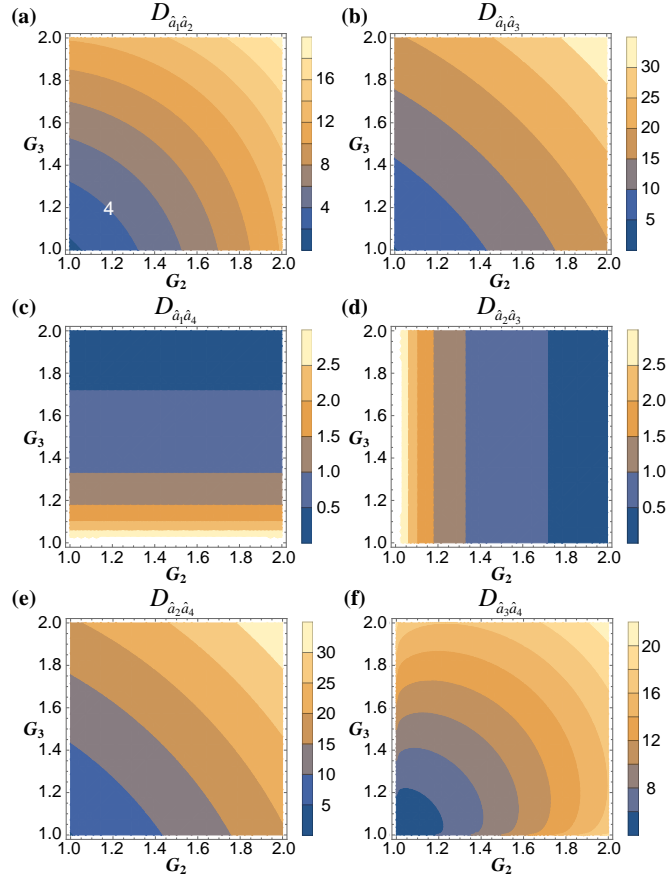

Figure 5. Bipartite entanglement for the four-mode DELC-FWM processes using Duan criterion. (a)-(f) Contour plot of $D_{12}$, $D_{13}$, $D_{14}$, $D_{23}$, $D_{24}$ and $D_{34}$ with different values of gains $G_2$ and $G_3$ at $G_1 = 1.1$.

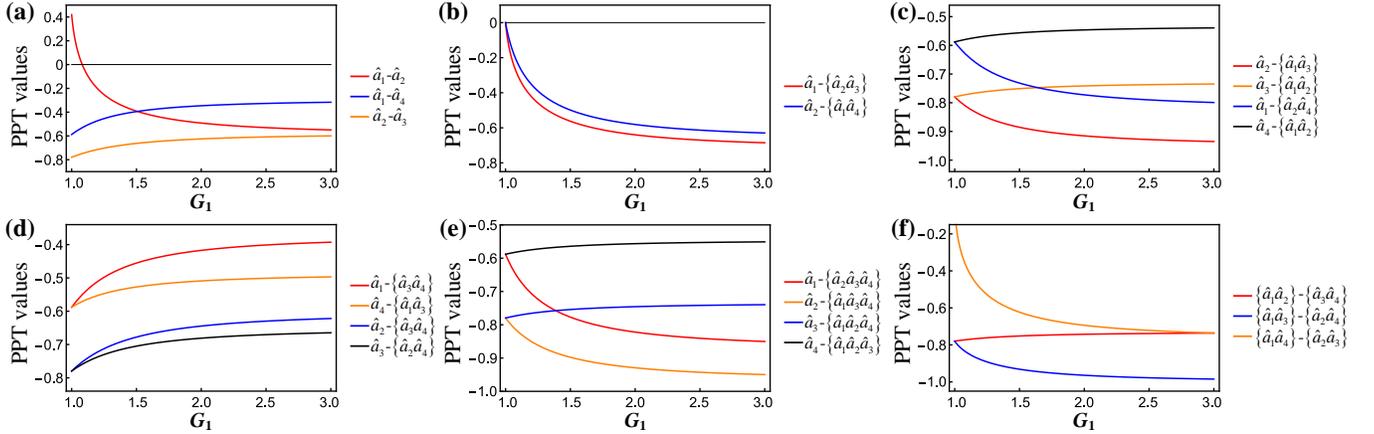

Figure 6. PPT values of the four-mode DELC-FWM processes with increasing $G_1$ at $G_2 = 1.3$ and $G_3 = 1.1$. (a) Bipartite entanglement. (b)-(d) Tripartite entanglement. (d)(e) Quadripartite entanglement (two types of 1-3 and 2-2 modes).

The dependence of PPT criterion on gain $G_1$ in the four-mode system are shown in Figure 6. Here we set the values $G_2 = 1.3$ and $G_3 = 1.1$. The PPT values of $\hat{a}_1$-$\hat{a}_3$, $\hat{a}_2$-$\hat{a}_4$ and $\hat{a}_3$-$\hat{a}_4$ are not negative when $G_1 > 1$ meaning the absence of entanglement, which are similar as the Duan criterion. The modes $\hat{a}_1$ and $\hat{a}_2$ do not entangled when $G_1$ is small, because the interaction of these two modes is not strong enough in the Hamitonian in Equation 6*a*. The entanglement of the three-mode subsystem in the four-mode system is all verified in Figure 6(b)-(d). $\hat{a}_3$-$\{\hat{a}_1\hat{a}_4\}$ and $\hat{a}_4$-$\{\hat{a}_2\hat{a}_3\}$ do not entangled and therefore do not given in this figure. This is because mode $\hat{a}_3$ ($\hat{a}_4$) do not interact with any one mode of modes $\{\hat{a}_1\hat{a}_4\}$ ($\{\hat{a}_2\hat{a}_3\}$) in this symmetrical structure. The PPT values of 1-3 modes and 2-2 modes are all less than 0 when $G_1 > 1$. However, increasing $G_1$ slightly decrease the entanglement of $\{\hat{a}_1\hat{a}_2\}$-$\{\hat{a}_3\hat{a}_4\}$, $\hat{a}_3$-$\{\hat{a}_1\hat{a}_2\hat{a}_4\}$ and $\hat{a}_4$-$\{\hat{a}_1\hat{a}_2\hat{a}_3\}$.

## 4. Coherent Channels with Dressing Effect

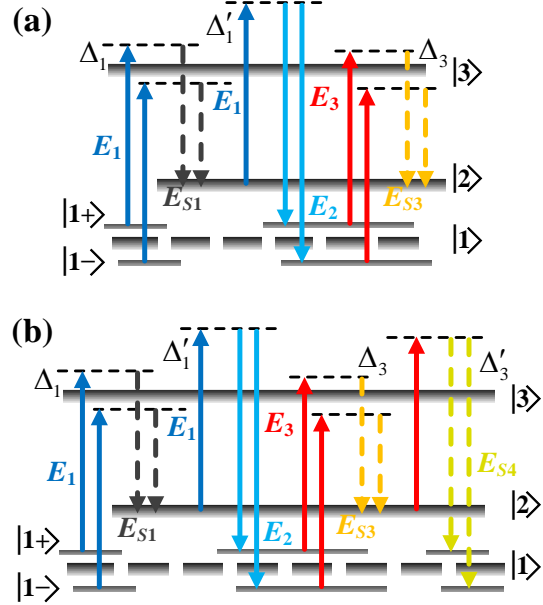

Figure 7. The energy-level diagram in the dressed-state picture. (a) Three-mode system. (b) Four-mode system. The energy levels $|1\pm\rangle$ are created from $|1\rangle$. Multiple coherent channels of FWM are constructed because of the strong dressing effect of the field $\boldsymbol{E}_1$.

In this section, we turn our attention to the atomic coherence control. Here, we take the three-mode system as an example (Figure 7). A laser field, specially from the same laser with pump field $\boldsymbol{E}_1$, is added as a dressing field to modulate the energy levels involved in the PA-FWM process. The derivative process of three output signals $\boldsymbol{E}_{S1}$, $\boldsymbol{E}_{S2}$, and $\boldsymbol{E}_{S3}$ can be described by the perturbation chains:

$$\text{PA-FWM1} \quad \rho_{22}^{(0)} \xrightarrow{\omega_1} \rho_{32}^{(1)} \xrightarrow{\omega_2} \rho_{12}^{(2)} \xrightarrow{\omega_1} \rho_{32(S1)}^{(3)}$$

$$\text{PA-FWM1} \quad \rho_{11}^{(0)} \xrightarrow{\omega_1} \rho_{31}^{(1)} \xrightarrow{\omega_{S1}} \rho_{21}^{(2)} \xrightarrow{\omega_1} \rho_{31(S2)}^{(3)}$$

$$\text{PA-FWM2} \quad \rho_{11}^{(0)} \xrightarrow{\omega_3} \rho_{31}^{(1)} \xrightarrow{\omega_{S3}} \rho_{21}^{(2)} \xrightarrow{\omega_1} \rho_{31(S2)}^{\prime(3)}$$

$$\text{PA-FWM2} \quad \rho_{22}^{(0)} \xrightarrow{\omega_1} \rho_{32}^{(1)} \xrightarrow{\omega_2} \rho_{12}^{(2)} \xrightarrow{\omega_3} \rho_{32(S3)}^{(3)}$$

where $\rho$ is the density matrix element and the superscripts (0), (1), (2), or (3) express the perturbation order. In the dressed-state pictures, the third-order nonlinear susceptibility $\chi^{(3)}$ is contained in $\rho^{(3)}$, which can be expressed as follows according to the perturbation chains:

$$\text{PA-FWM1} \quad \rho_{32(S1)}^{(3)} = -i\Omega_1^2\Omega_2/d_{32}d_{12}d_{32}' \tag{12a}$$

$$\text{PA-FWM1} \quad \rho_{31(S2)}^{(3)} = -i\Omega_1^2\Omega_{S1}/d_{31}d_{21}d_{31}' \tag{12b}$$

$$\text{PA-FWM2} \quad \rho_{31(S2)}^{\prime(3)} = -i\Omega_1\Omega_3\Omega_{S3}/d_{31}''d_{21}'d_{31}''' \tag{12c}$$

$$\text{PA-FWM2} \quad \rho_{32(S3)}^{(3)} = -i\Omega_1\Omega_2\Omega_3/d_{32}''d_{12}'d_{32}''' \tag{12d}$$

Table 1. Resonance frequencies of the coherent channels in the three-mode DELC-FWM processes.

| Coherence Channels | Resonance Frequencies of the Coherent Channels | | | |
|---|---|---|---|---|
| | $\delta_1$ | $\delta_2$ | $\delta_2'$ | $\delta_3$ |
| C1 | $\delta_1 = -\Delta_1'$ | $\delta_2 = \Delta_1'$ | $\delta_2' = \Delta_1'$ | $\delta_3 = -\Delta_1'$ |
| C2 | $\delta_1 = \dfrac{\Delta_1 + \sqrt{\Delta_1^2 + 4\Gamma_{21}\Gamma_{23} + 4\Omega_1^2}}{2}$ | $\delta_2 = -\dfrac{\Delta_1 + \sqrt{\Delta_1^2 + 4\Gamma_{21}\Gamma_{23} + 4\Omega_1^2}}{2}$ | $\delta_2' = -\dfrac{\Delta_1 + \sqrt{\Delta_1^2 + 4\Gamma_{21}\Gamma_{23} + 4\Omega_1^2}}{2}$ | $\delta_3 = \dfrac{\Delta_1 + \sqrt{\Delta_1^2 + 4\Gamma_{21}\Gamma_{23} + 4\Omega_1^2}}{2}$ |
| C3 | $\delta_1 = \dfrac{\Delta_1 - \sqrt{\Delta_1^2 + 4\Gamma_{21}\Gamma_{23} + 4\Omega_1^2}}{2}$ | $\delta_2 = -\dfrac{\Delta_1 - \sqrt{\Delta_1^2 + 4\Gamma_{21}\Gamma_{23} + 4\Omega_1^2}}{2}$ | $\delta_2' = -\dfrac{\Delta_1 - \sqrt{\Delta_1^2 + 4\Gamma_{21}\Gamma_{23} + 4\Omega_1^2}}{2}$ | $\delta_3 = \dfrac{\Delta_1 - \sqrt{\Delta_1^2 + 4\Gamma_{21}\Gamma_{23} + 4\Omega_1^2}}{2}$ |

where $\Omega_i$ is the Rabi frequency of field $E_i$. $d_{32} = \Gamma_{32} + i\Delta_1'$, $d_{12} = \Gamma_{12} + i(\Delta_1' - \Delta_{S2})$, $d_{32}' = \Gamma_{32} + i(\Delta_1' - \Delta_{S2} + \Delta_1)$, $d_{31} = \Gamma_{31} + i\Delta_1$, $d_{21} = \Gamma_{21} + i(\Delta_1 - \Delta_{S1})$, $d_{31}' = \Gamma_{31} + i(\Delta_1 - \Delta_{S1} + \Delta_1')$, $d_{31}'' = \Gamma_{31} + i\Delta_3$, $d_{21}' = \Gamma_{21} + i(\Delta_3 - \Delta_{S3})$, $d_{31}''' = \Gamma_{31} + i(\Delta_3 - \Delta_{S3} + \Delta_1')$, $d_{32}'' = \Gamma_{32} + i\Delta_1'$, $d_{12}' = \Gamma_{12} + i(\Delta_1' - \Delta_{S2}')$, $d_{32}''' = \Gamma_{32} + i(\Delta_1' - \Delta_{S2}' + \Delta_3)$. $\Delta_{S1}$, $\Delta_{S2}$ and $\Delta_{S3}$ represent the frequency detuning of the signals $E_{S1}$, $E_{S2}$ and $E_{S3}$. $\Delta_1$ and $\Delta_1'$ are the frequency detuning of the fields $E_1$ from the transitions $|1\rangle \to |3\rangle$ and $|2\rangle \to |3\rangle$, respectively, defined as the difference between the resonant transition frequency and laser frequency; $\Delta_3$ is the frequency detuning of the fields $E_3$. $\Gamma_{ij} = (\Gamma_i + \Gamma_j)/2$ is the transverse relaxation rate between states $|i\rangle$ and $|j\rangle$.

Due to the frequencies $\omega_{Si}$ of generated photons with small quantum deviations $\delta_i$ around the corresponding central frequency $\varpi_{Si}$, $\omega_{Si}$ can be written as $\omega_{Si} = \varpi_{Si} + \delta_i$ ($i = 1, 2, 3$) with the limitation of $|\delta_i| \ll \varpi_{Si}$. Considering quantum deviation $\delta_1$, the frequency detuning of the signals $E_{S1}$ can be expressed as $\Delta_{S1} = \omega_{32} - \omega_{S1} = \omega_{32} - (\varpi_{S1} + \delta_1) = \Delta_1 - \delta_1$. According to the conservation of energy in PA-FWM1 and in PA-FWM2, $\delta_2 = -\delta_1$ and $\delta_3 = \delta_1$, which show the frequency correlation of the triphoton state in this three-mode energy-level-cascaded system. Different quantum properties of triphoton state also can be reflected by focusing on these photon deviations. In the resonance conditions of coupling fields, the third-order density matrix element $\rho_{31(S2)}^{(3)}$ in Equation 12b with the expression of $\Delta_{S1}$ can be rewritten as

$$\rho_{31(S2)}^{(3)} = \frac{-i\Omega_1^2 \Omega_{S1}}{(\Gamma_{31} + i\Delta_1)(\Gamma_{21} + i\delta_1)(\Gamma_{31} + i\delta_1 + i\Delta_1')} \tag{13}$$

Using atomic coherence, parametric gain properties of this quantum interplay can be directly adjusted by modulating the dressing effect. When we consider the dressing effect of $E_1$, the perturbation chain of $E_{S2}$ can be expressed as $\rho_{11}^{(0)} \xrightarrow{\omega_1} \rho_{31}^{(1)} \xrightarrow{\omega_{S1}} \rho_{2\Omega_1\pm}^{(2)} \xrightarrow{\omega_1} \rho_{31(S2)D}^{(3)}$. The subscript "1" of $\rho_{21}^{(2)}$ is replaced by "$G_1\pm$", which indicates that dressing fields $E_1$ dress the level $|1\rangle$ and influence the identical coherence between states $|1\rangle$ and $|3\rangle$. The third-order density matrix element in Equation 13 with single dress can be rewritten as follows:

$$\rho_{31(S2)D}^{(3)} = \frac{-i\Omega_1^2 \Omega_{S1}}{\left[ (\Gamma_{31} + i\Delta_1)\left(\Gamma_{21} + i\delta_1 + \dfrac{\Omega_1^2}{\Gamma_{23} + i\delta_1 - i\Delta_1}\right) \right] (\Gamma_{31} + i\delta_1 + i\Delta_1')} \tag{14}$$

By maximizing the denominator of the third-order density matrix element, we can obtain the resonance positions as shown in Table 1. Figure 8 shows the third-order density matrix element in the perturbation chains under different conditions and exhibits the resonance positions of small quantum deviations window $\delta_i$, which is the frequency corresponding to the coherent channels. The first row and the second row are the Stokes signal and anti-Stokes signal of the PA-FWM1, respectively. The third row and the fourth row are the Stokes signal and anti-Stokes signal of the PA-FWM2, respectively. A set of longitudinal resonance positions corresponds to one coherent channel, and also a single FWM mode. The dressing field $E_1$ creates the dressed states $|1\pm\rangle$ from $|1\rangle$ as shown in Figure 7. It results in two PA-FWMs coherent channels coexisted, and the output signals thus have two modes of different frequencies and therefore the number of coherent channels are increased. In figure 8, with the dressing effect, peak (II) split into two peaks (IV) and (V) denoted by the dressing term in Equation 14. Three coherent channels coexist (IV, V, and VI), and all satisfying the energy conservation condition $\delta_1 + \delta_2 + \delta_2' + \delta_3 = 0$. The quantum information capacity, expressed as $n^3$ where $n$ represents the number of coherent channels, is greatly expanded to $3^3$ in this single dressing system.

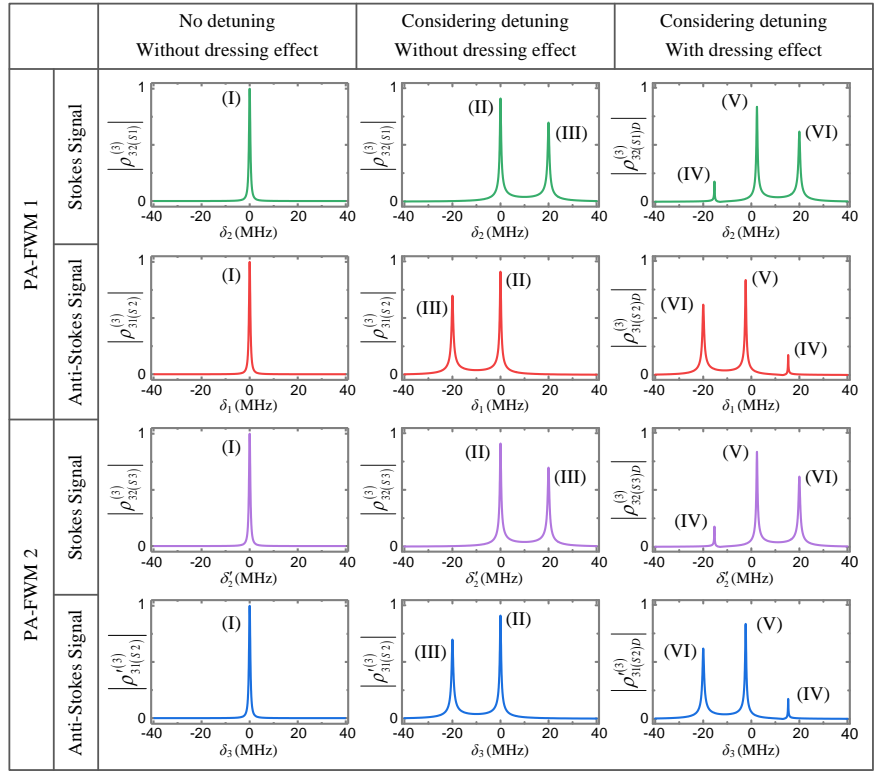

Figure 8. Theoretically calculated normalized third-order density matrix where the dressing effect acts on the second-order density matrix element $\rho_{21}^{(2)}$. The conditions of three columns are different. The conditions of the first column are $\Delta_1 = \Delta_1' = 0$, without the dressing effect. The conditions of the second column are $\Delta_1 = 13$ MHz and $\Delta_1' = 20$ MHz, without the dressing effect. The conditions of the third column are $\Delta_1 = 13$ MHz and $\Delta_1' = 20$ MHz, with the dressing effect of field $E_1$.

Subsequently, Duan and PPT criteria in each coherent channel are investigated in the three- and four-mode DELC-FWM system. According to Equation 14, we can obtain the dressing-modulated optical gain, and then the entanglement characteristics can be actively and directly controlled in the process of preparing the entangled sources. Figure 9 shows the values of criteria versus $\delta_1$, where the gain of the first PA-FWM modulated by the dressing field $E_1$. Here we set $G_2 = 1.2$ in three-mode system and $G_2 = 1.3$, $G_3 = 1.1$ in four-mode system. The red areas, where the Duan are not violated or the PPT values are not negative, indicate the absence of entanglement. The unentangled modes are not given in this figure, including $\hat{a}_1 - \hat{a}_3$ in three-mode and $\hat{a}_1$-$\hat{a}_3$, $\hat{a}_2$-$\hat{a}_4$, $\hat{a}_3$-$\hat{a}_4$, $\hat{a}_3$-$\{\hat{a}_1\hat{a}_4\}$, and $\hat{a}_4$-$\{\hat{a}_2\hat{a}_3\}$ in four-mode. The black dashed areas correspond to the coherent channels, where the gains are stronger Also, the gains in channels are independent and do not interfere with each other, so the values of the criterion are different.

## 5. Discussion and Conclusion

In addition to the case of the second-order density matrix element $\rho_{21}^{(2)}$ dressed by the field $E_1$ mentioned above, it can also be expanded to many other dressing cases. The first-order density matrix element $\rho_{31}^{(1)}$ under the contidion of the energy level $|1\rangle$ dressed by the field $E_1$ can be written as Equation 15, which is obtained by the perturbation chain $\rho_{11}^{(0)} \xrightarrow{\omega_1} \rho_{3\Omega_1\pm}^{(1)} \xrightarrow{\omega_{S1}} \rho_{21}^{(2)} \xrightarrow{\omega_1} \rho_{31(S2)D}^{(3)}$.

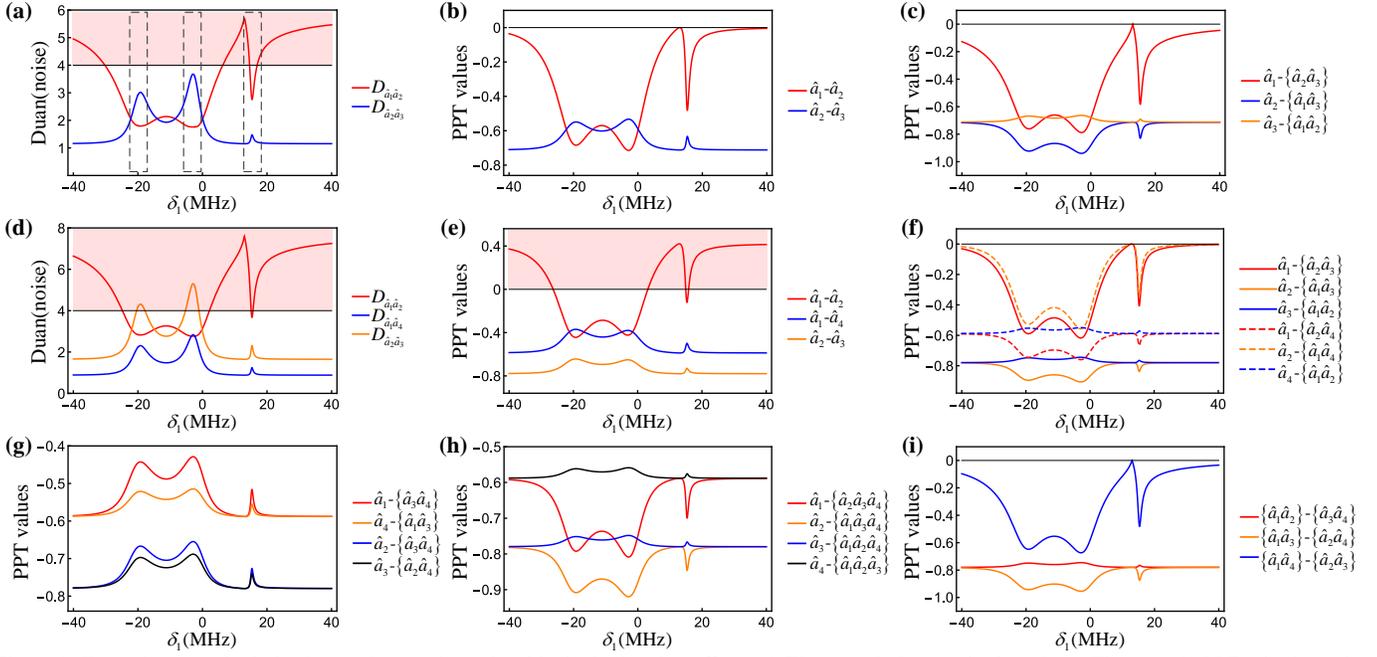

Figure 9. Entanglement criteria in three coherent channels with single dressing effect. (a) Three-mode Duan criterion. (b)(c) Three-mode PPT criterion. (d) Four-mode Duan criterion. (e)-(i) Four-mode PPT criterion. The red areas indicate no entanglement. The black dashed areas show the criterion in each channel.

$$\rho^{(3)}_{31(S2)D} = \frac{-i\Omega_1^2 \Omega_{S1}}{\left(\Gamma_{31} + i\Delta_1 + \dfrac{\Omega_1^2}{\Gamma_{33}}\right)(\Gamma_{21} + i\delta_1)(\Gamma_{31} + i\delta_1 + i\Delta_1')} \tag{15}$$

The resonance positions of the coherent channels are $\delta_1 = 0$ and $\delta_1 = -\Delta_1'$. The number of coherent channels is still 2, and the resonance peaks do not splitted by the dressing effect, which is shown in Figure A3 in Appendix. The third-order density matrix element $\rho^{(3)}_{31}$ dressed by the field $E_1$ at energy level $|1\rangle$ can be written as Equation 16, which is obtained by the perturbation chain $\rho^{(0)}_{11} \xrightarrow{\omega_1} \rho^{(1)}_{31} \xrightarrow{\omega_{S1}} \rho^{(2)}_{21} \xrightarrow{\omega_1} \rho^{(3)}_{3\Omega_1 \pm (S2)D}$.

$$\rho^{(3)}_{31(S2)D} = \frac{-i\Omega_1^2 \Omega_{S1}}{\left[\begin{array}{c}(\Gamma_{31} + i\Delta_1)(\Gamma_{21} + i\delta_1) \\ \left(\Gamma_{31} + i\delta_1 + i\Delta_1' + \dfrac{\Omega_1^2}{\Gamma_{33} + i\delta_1 + i\Delta_1' - i\Delta_1}\right)\end{array}\right]} \tag{16}$$

The resonance positions of the three coherent channels are $\delta_1 = 0$, $\delta_1 = \left(\Delta_1 - 2\Delta_1' + \sqrt{\Delta}\right)/2$ and $\delta_1 = \left(\Delta_1 - 2\Delta_1' - \sqrt{\Delta}\right)/2$, where $\Delta = (\Delta_1 - 2\Delta_1')^2 - 4(\Delta_1'^2 - \Delta_1\Delta_1' - \Omega_1^2 - \Gamma_{31}\Gamma_{33})$, which is similar to Figure 8 but the frequency difference of the latter two peaks is $\sqrt{\Delta}$. The phenomenon of splitted resonance peaks observed in the experiment is generally caused by the dressed second-order density matrix element $\rho^{(2)}$, because the detuning of deriving $\rho^{(2)}$ is smaller, with stronger dressing effect, compared with deriving $\rho^{(3)}$.

Moreover, the pump field $E_3$ can also be used as a dressing field. The second-order density matrix element $\rho^{(2)}_{21}$ dressed by the field $E_3$ at energy level $|1\rangle$ can be written as Equation 17, which is obtained by the perturbation chain $\rho^{(0)}_{11} \xrightarrow{\omega_1} \rho^{(1)}_{31} \xrightarrow{\omega_{S1}} \rho^{(2)}_{2\Omega_3 \pm} \xrightarrow{\omega_1} \rho^{(3)}_{31(S2)D}$.

$$\rho_{31(S2)D}^{(3)} = \frac{-i\Omega_1^2\Omega_{S1}}{\begin{bmatrix}(\Gamma_{31}+i\Delta_1)\left(\Gamma_{21}+i\delta_1+\dfrac{\Omega_3^2}{\Gamma_{23}+i\delta_1-i\Delta_3}\right)\\(\Gamma_{31}+i\delta_1+i\Delta_1')\end{bmatrix}} \quad (17)$$

Here the resonance positions and the optical gains of the coherent channels depends on the Rabi frequency $\Omega_3$ and frequency detuning $\Delta_3$ of the dressing field $E_3$. The three coherent channels are at $\delta_1 = \left(\Delta_3 + \sqrt{\Delta_3^2 + 4\Gamma_{21}\Gamma_{23} + 4\Omega_3^2}\right)/2$, $\delta_1 = \left(\Delta_3 - \sqrt{\Delta_3^2 + 4\Gamma_{21}\Gamma_{23} + 4\Omega_3^2}\right)/2$ and $\delta_1 = -\Delta_1'$.

In summary, the DELC-FWM processes are proposed to one-step produce the three- and four-mode quantum entangled states in Rb atomic vapors. We apply Duan and PPT criteria to characterize the multipartite entanglement potentially existed in this cascaded system and theoretically investigate the dependence of criteria on the system parameters. Furthermore, the dressing field is introduced to produce and coherent control the multimode multiplexed entanglement via constructing multiple coherent channels of FWM. The properties of the entanglement among output beams is coherent-channel dependent, and can be well controlled using the dressing effect of atoms and many optical parameters in our system, without need of extra control (e.g., beam splitters) after the quantum interplay. In our scheme, using atomic coherence, the generation and modulation of multipartite entanglement can be integrated in the process of the entangled states preparation. These results may be helpful for multimode quantum secure communication, quantum routing and quantum coherent control.

## 6. Appendix

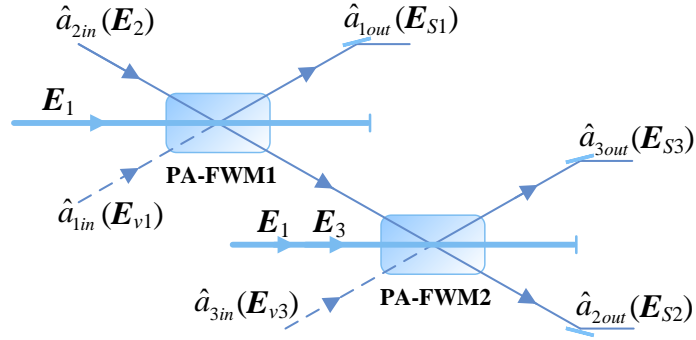

Figure A1. Schematic diagram of cascading two Rb cells to generate three-mode entangled states. $\hat{a}_{2in}$ is the seed input signal; $\hat{a}_{1in}$ and $\hat{a}_{3in}$ are the vacuum input signals. $E_2$ is amplified by PA-FWM1 and injected into PA-FWM2. $\hat{a}_{1out}$, $\hat{a}_{2out}$ and $\hat{a}_{3out}$ are three output signals.

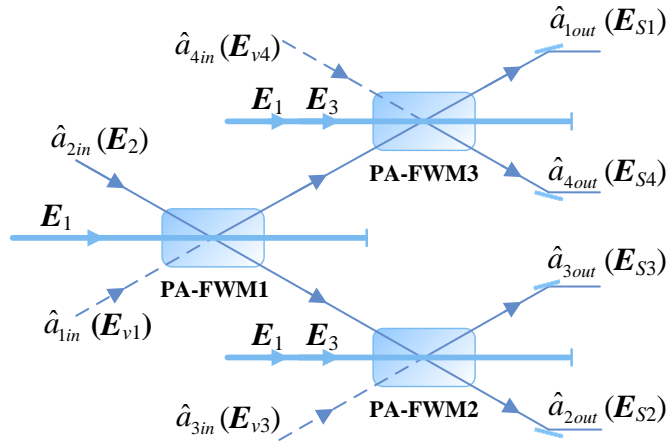

Figure A2. Schematic diagram of cascading three Rb cells to generate four-mode entangled states. $\hat{a}_{2in}$ is the seed input signal; $\hat{a}_{1in}$, $\hat{a}_{3in}$ and $\hat{a}_{4in}$ are the vacuum input signals. The amplified $E_2$ and amplified $E_{v1}$ via PA-FWM1 process are injected into PA-FWM2 and PA-FWM3, respectively. $\hat{a}_{1out}$, $\hat{a}_{2out}$, $\hat{a}_{3out}$ and $\hat{a}_{4out}$ are four output signals.

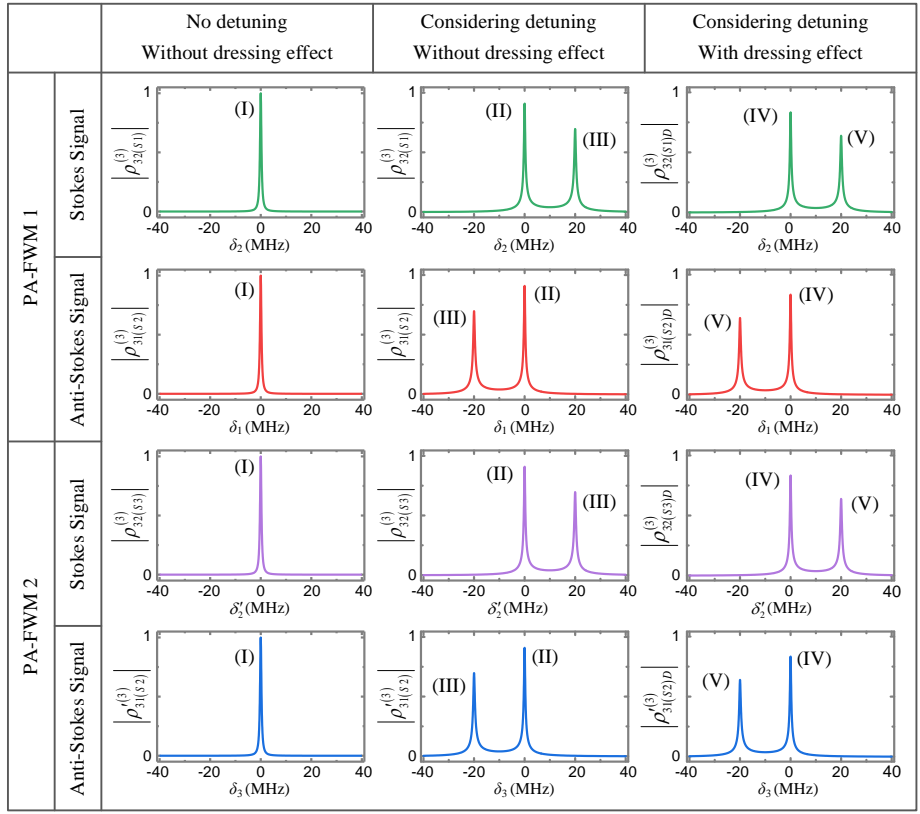

Figure A3. Theoretically calculated normalized third-order density matrix element where the dressing effect acts on the first-order density matrix element $\rho_{31}^{(1)}$. The conditions of three columns are different. The conditions of the first column are $\Delta_1 = \Delta_1' = 0$, without the dressing effect. The conditions of the second column are $\Delta_1 = 13$ MHz and $\Delta_1' = 20$ MHz, without the dressing effect. The conditions of the third column are $\Delta_1 = 13$ MHz and $\Delta_1' = 20$ MHz, with the dressing effect of field $E_1$.

The three- and four-mode DELC-FWM systems correspond to cascade Rb cells in terms of obtaining multiple entangled states as shown in Figure A1 and A2. As DELC-FWM only requires a single Rb cell, it introduces fewer vacuum losses. Moreover, this method is phase insensitive without the need of complicated phase locking technique.

Figure A3 shows the resonance positions of small quantum deviations window $\delta_i$ under the condition of the dressed first-order density matrix element $\rho^{(1)}$ (Equation 15). Since the process of deriving $\rho^{(1)}$ is independent of the quantum deviation, the resonance peak does not split because of the dressing effect acting on $\rho^{(1)}$.

## ACKNOWLEDGMENTS

This work was supported by the National Key R&D Program of China (2017YFA0303700, 2018YFA0307500), Key Scientific and Technological Innovation Team of Shaanxi Province (2021TD-56), National Natural Science Foundation of China (61975159, 11904279, 12174302, 62022066, 12074306, 12074303).